\documentclass[]{cupconf}
\usepackage{graphicx}

\title[Sulphur Abundances in Disk Stars from the forbidden SI Line at 10821 \AA\,]
{Sulphur Abundances in Disk Stars from the Forbidden SI Line at
10821 \AA\ }

\author[N. Ryde]{\\ Nils Ryde$^1$}
\affiliation{$^1$ Department of Astronomy and Space Physics, Box
515, SE-751 20 Uppsala, Sweden }

\begin{document}
\maketitle

\begin{abstract}

\end{abstract}

\firstsection
\section{Context} 
\begin{figure}
\includegraphics[width=\textwidth]{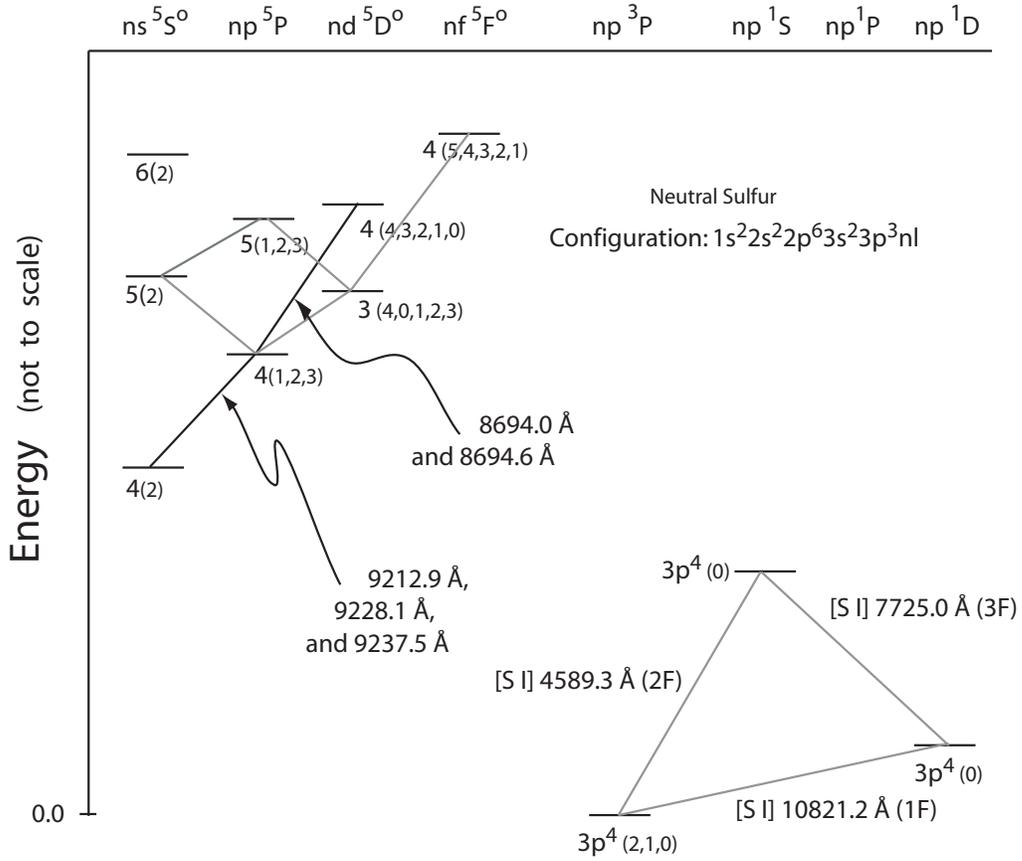}
\caption{Grotrian diagram of sulphur indicating spectral
abundance-tracers which are possible to use. }\label{fig1}
\end{figure}

In this poster, we present an investigation of a new, preferred diagnostic
tool for the determination of the sulphur abundance of disk stars,
see also Ryde A\&A (2006). We are in the process of analyzing a
large set of stars both in the galactic disk and halo (Ryde et al.
2007, in prep.). This diagnostics, the forbidden sulphur line at
10821 \AA\  (first observed in the Sun by Swings et al. 1968), is
less sensitive to the assumed temperatures of the stars investigated
and less prone to non-LTE effects than other tracers. In the
Grotrian diagram (Figure \ref{fig1}), we have indicated the most
commonly used tracers of stellar sulphur abundances, and three
forbidden transitions. The forbidden line with the predicted largest
strength is the $^3P_2-^1D_2$ transition at 10821 \AA . It is an
intercombination line, an M1 transition between the triplet
ground-state and the first excited singlet state. The E2 transition
is two orders of magnitude weaker. In the investigation presented
here, we have studied a homogeneous set of sub-giant and giant stars
ranging from spectral types of G5 to K4 (effective temperatures of
$4000- 5000$ K). The reason why the [SI] has not been use before is
that it lies beyond the reach of normal CCDs.

\begin{figure}
\includegraphics[width=\textwidth]{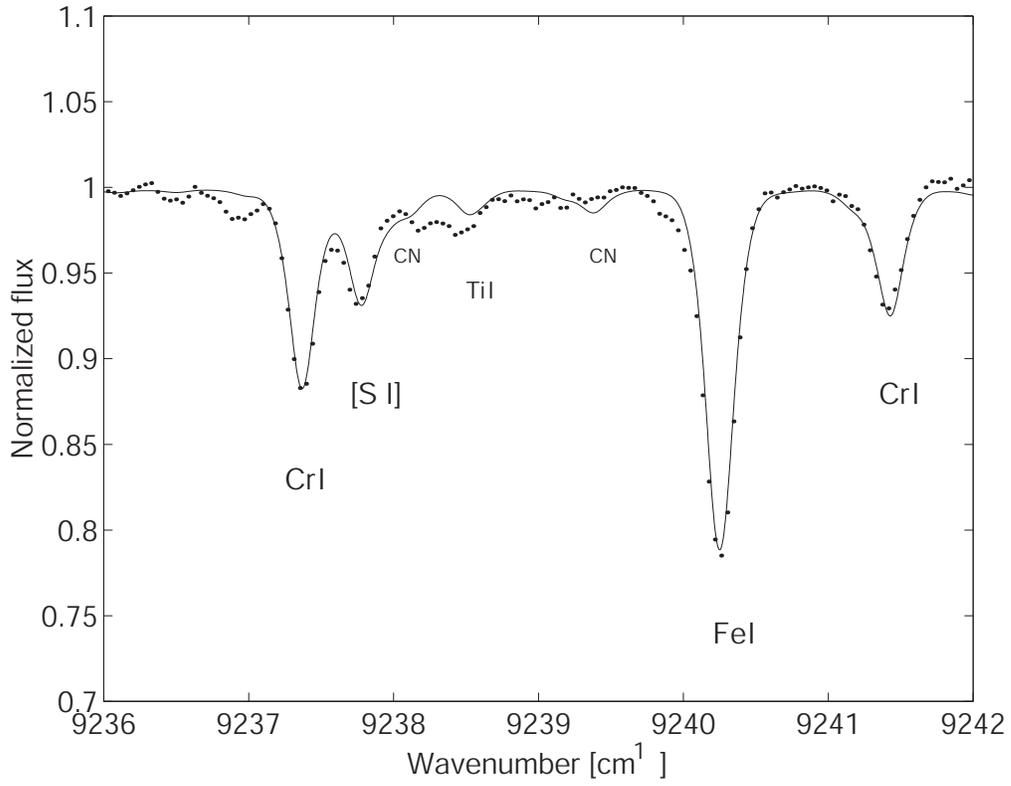}
\caption{ The spectrum of HD212943 is shown as by dots.  A full line
shows a synthetic spectrum with [S/Fe] = 0.21. }\label{fig2}
\end{figure}

\begin{figure}
\includegraphics[width=\textwidth]{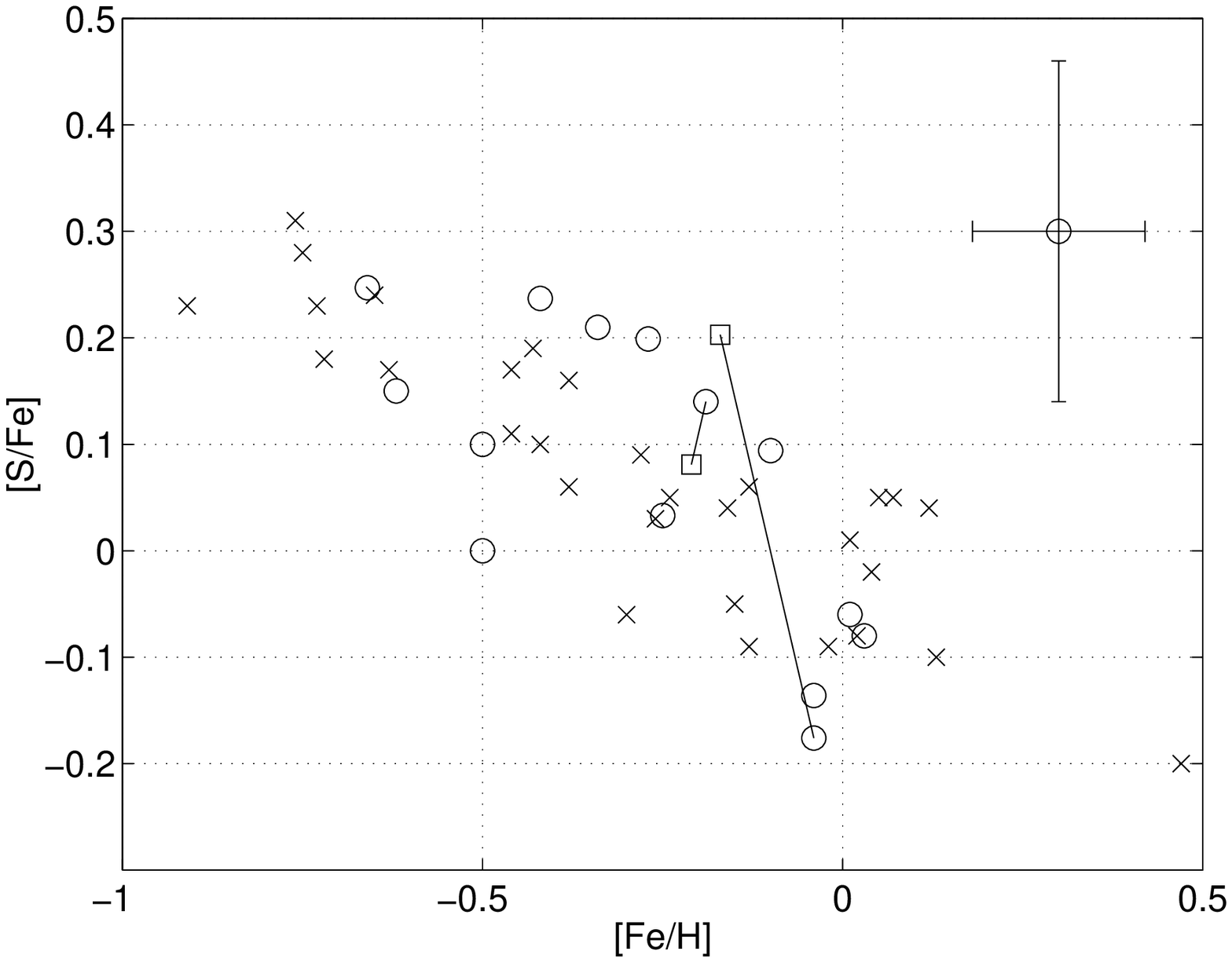}
\caption{The [S/Fe] vs [Fe/H] plot, in which we also have plotted
the determinations of Chen et al. (2002) by crosses. For details see
Ryde (2006).}\label{fig3}
\end{figure}

\section{Method}
High-resolution, near infrared spectra of the [SI] line are recorded
using the Phoenix spectrometer on the Gemini South telescope. Our
high signal-to-noise spectra have a resolution of $R = 60,000$. We
have modelled our observed spectra with synthetic spectra based on
MARCS model atmospheres. The stellar parameters of the stars are
taken from the Elodie library (Soubiran et al. 1998), and the line
strength of the [SI] line is taken from the NIST database. In Figure
\ref{fig2}, we show an example of a spectrum of the [SI] line. The
star, HD212943, has a metallicity of [Fe/H] = -0.34, a temperature
of 4590 K, and log(g) = 2.8.

\section{Results}  We show that the [SI] line at 10821 \AA\  is detectable and
useful for an analysis of sulphur abundances in disk stars. Based on
this line we can corroborate the alpha element-behaviour of sulphur
in the disk, which can be seen in the [S/Fe] vs [Fe/H] plot in
Figure \ref{fig3} and from the discussion in Ryde (2006) and Chen et
al. (2002).


\begin{thebibliography}{99}

\bibitem[]{Chen2002}
         Chen et al. (2002).
         \textit{A\&A} \textbf{390}, 225.

\bibitem[]{ryde2006}
    Ryde (2006).
    \textit{A\&A} \textbf{455}, L13.

\bibitem[]{Sou98}
         Soubiran et al. (1998).
         \textit{A\&A Suppl.} \textbf{133}, 221.

\bibitem[]{Swings68}
         Swings et al. (1968).
         \textit{SoPh} \textbf{6}, 3.

\end{thebibliography}
\end{document}